\documentstyle[12pt]{article}    
\begin{document}
\title{Boost invariant quantum evolution of a 
meson field at large proper times \\
}
\author{
Dominique Vautherin \\ 
{\normalsize 
Division de Physique Th\'eorique \thanks{Unit\'e de Recherche des 
Universit\'es 
Paris XI et Paris VI associ\'ee au C.N.R.S}} \\
{\normalsize  Institut de Physique Nucl\'eaire,} \\
{\normalsize F-91406 , Orsay Cedex, France } \\
~~ \\
and \\
~~\\
T. Matsui \\
{\normalsize Yukawa Institute for Theoretical Physics} \\
{\normalsize Kyoto University, Kyoto 606, Japan} \\
}

\date{September 1996 Preprint IPNO/TH 96-31 and YITP-96-39}
\maketitle
\begin{abstract}
We construct asymptotic solutions of the functional Schr\"odinger equation
for a scalar field in the Gaussian approximation at large proper time. These
solutions describe the late proper time stages of the expansion of a meson
gas with boost invariant boundary conditions. The relevance of these
solutions for the formation of a disoriented chiral condensate in ultra
relativistic collisions is discussed. 
\end{abstract}

\newpage

\paragraph*{Introduction. ---}
\label{sect_introduction}

Several recent studies have been devoted to the calculation of the
evolution of a self interacting scalar field.  As an example the
evolution of a scalar field is an important ingredient in most
inflationary models \cite{GUTH,JACKIW,DEVEGA,SAMIULLAH,MOTTOLA}.  Also
classical as well as quantum calculations of the evolution of a scalar
field have been considered in the context of the possible formation of
a disoriented chiral condensate in ultrarelativistic nuclear
collisions \cite{BLAIZOT,WILCZEK,HOLMAN,SINGH,ANSELM}.

In this note we investigate the asymptotic forms of the
quasi-one-dimensional quantum evolution of the meson fields which
possess a symmetry with respect to the Lorentz boost along
$z$-direction. Such a symmetry was originally introduced by
Bjorken\cite{BJORKEN} in a hydrodynamical model of ultrarelativistic
nucleus-nucleus collision based on the physical assumption of central
plateau formation in the final particle distribution in the rapidity
space.

We take a simple model for mesons described by a self-interacting
scalar field $\varphi({\bf x})$ and adopt the functional Schr\"odinger
picture with the Hamiltonian density: 
\begin{displaymath}
{\cal H}({\bf x}) = \frac{1}{2} \pi^2 ({\bf x}) + \frac{1}{2} (\vec \nabla 
\varphi ({\bf x}) )^2 
+ \frac{m_0^2}{2} \varphi^2 ({\bf x}) + \frac{\lambda}{24} 
\varphi^4 ({\bf x}) \ . 
\end{displaymath}
We use the framework of the Gaussian or mean field approximation, i.e.
we assume that at each time $t$ the wave functional describing the
system is a gaussian type coherent state
\begin{eqnarray*} 
\Psi \left[ \varphi ({\bf x}) \right] &=& {\cal N} 
\exp \left( i< \pi | \varphi - \bar \varphi> \right) \\
&& \times \exp \left( -< \varphi- \bar \varphi | ( \frac{1}{4G} + i \Sigma ) | 
\varphi- \bar \varphi > \right) \mbox{,}
\end{eqnarray*}
where $G$, $\Sigma$, $\bar \varphi$, $\pi$ define respectively the
real and imaginary part of the kernel of the Gaussian and its average
position and momentum.  We impose Lorentz invariant boundary conditions
i.e. we assume that observables depend only on the proper time
variable $\tau= \sqrt{t^2 - z^2}$.  We examine the behaviour of the
meson wave functional for large values of the proper time variable. We
will show that this limit can be worked out analytically by solving
the linearized evolution equations.

For more simplicity we restrict ourselves to the case of a single
scalar field with an initial condition which reduces to a single pure
state (not a statistical mixture). As will be shown below this case is
already instructive and general enough to provide a transparent
interpretation of some of the numerical calculations performed
recently by Cooper {\it et al} \cite{COOPER} in the more general case
of a sigma model with an initial state which is a statistical mixture. 
A more specific and more detailed comparison adapted to this case will
be presented in a forthcoming publication.

\paragraph*{Mean Field Equations. ---}
\label{sect_mf}

The evolution equations in the Gaussian approximation read
\cite{JACKIW,SAMIULLAH,MOTTOLA}
\begin{eqnarray*}
\dot G&=& 2 (G \Sigma +\Sigma G), \\
&& \\
\dot \Sigma&=& \frac{1}{8} G^{-2} - 2\Sigma^2 -
\frac{1}{2} \{ -\Delta + m_0^2 + \frac{\lambda}{2} \bar \varphi^2 +
\frac{\lambda}{2}  G({\bf x},{\bf x}) \}
\\
&& \\
\dot {\bar \varphi }&=& -\bar \pi ,
\\
&& \\
\dot {\bar \pi}&=& 
\{ -\Delta + m_0^2 + \frac{\lambda}{6} \bar \varphi^2 +
\frac{\lambda}{2}  G({\bf x},{\bf x}) \} \bar \varphi .
\end{eqnarray*}
In these equations $m_0$ is the bare mass of the scalar field and $\lambda$
the bare strength of the self coupling term which we write as $\lambda
\varphi^4$ in the Lagrangian density.
The vacuum state corresponds to the static solution
\begin{displaymath}
\Sigma=0,~~~~{\bar \pi}=0,~~~~{\bar \varphi}=\varphi_0,
~~~G=1/ 2 \sqrt{-\Delta + \mu^2},
\end{displaymath}
where the quantity $\mu$ has to satisfy the so- called 
gap equation \cite{KERMAN}
\begin{equation}
\mu^2= m_0^2 + \frac{\lambda}{2} <{\bf x}| \frac{1}{2 \sqrt{-\Delta + \mu^2}} |
{\bf x}> + \frac{\lambda}{2} \varphi_0^2.
\label{GAP}
\end{equation}
The previous equation requires a regularization scheme such as a
discretization of the Laplacian operator on a lattice with a mesh size
$\Delta x= 1/\Lambda$ or a cutoff $\Lambda$ in momentum space. To make the
gap equation finite when the scale $\Lambda$ goes to $\infty$ a popular 
prescription is to send the bare coupling constant to zero
according to the formula \cite{LIN}
\begin{displaymath}
\frac{1}{2 \lambda_R}= \frac{1}{\lambda}+ \frac{1}{16 \pi^2} \log 
(\frac{2 \Lambda}{e \mu}).
\end{displaymath}
This prescription is however not free from difficulties, see
\cite{DEVEGA,KERMAN,LIN}.

The mean field evolution equations can be written in a more compact
form which has the further advantage of being manifestly covariant
\begin{eqnarray*}
&m&^2(x)= m_0^2 +\frac{\lambda}{2} {\bar \varphi}^2(x) + 
\frac{\lambda}{2} <x| S |x> \mbox{,} \\
&\{& \Box+ m_0^2 + \frac{\lambda}{6} \bar \varphi^2 +
\frac{\lambda}{2} <x| S |x> \} {\bar \varphi}=0  \mbox{,}
\end{eqnarray*}
where $x= (x_0, x_1, x_2, x_3)$ and where $S$ is the Feynman
propagator in the presence of an $x$ dependent mass
\begin{displaymath}
S= \frac{i}{\Box+m^2(x)+ i \varepsilon}.
\end{displaymath}
A proof of this equation can be constructed by using the functional methods of
Cornwall, Jackiw and Tomboulis \cite{CORNWALL}. The kernel $G$ of the
Gaussian wave functional is related to the propagator $S$ via the relation
\begin{displaymath}
<{\bf x}| G(t) |{\bf x}>= <x| S |x>.
\end{displaymath}

To obtain the asymptotic form of the state of the meson field at large
proper times we assume that the propagator $S$ is close to its vacuum
value.  In this case we linerize the equations of motion around the
vacuum value by writing
\begin{equation}
S= \frac{i}{\Box+ \mu^2 + \delta m^2(x) + i \varepsilon}= S_0 +
S_0~( \delta m^2)~ S_0 + \ldots
\label{PROPAGATOR}
\end{equation}
Introducing the Fourier decomposition
\begin{eqnarray*}
&\delta& m^2(x)= \int \frac{d^4 q}{(2 \pi)^4} \exp(i q \cdot x) \delta m^2(q),
\\
&\delta& \varphi (x)= {\bar \varphi}(x)- \varphi_0= 
\int \frac{d^4 q}{(2 \pi)^4} \exp(i q \cdot x) 
\delta \varphi (q),
\end{eqnarray*}
equation (\ref{PROPAGATOR}) leads to
\begin{displaymath}
<x| \delta S |x> = \int \frac{d^4 q}{(2 \pi)^4} \exp(i q \cdot x) 
\Pi_0(q^2) \delta m^2(q),
\end{displaymath}
where $\Pi_0(q^2)$ is the standard polarization tensor
\begin{eqnarray*}
\Pi_0(q^2)&=&- \int \frac{d^4 p}{(2 \pi)^4} S_0(p) S_0(p+q) \\
&\simeq& - \frac{1}{16 \pi^2} \{ \log (\frac{4 \Lambda^2}{e^2 \mu^2})
+ \frac{q^2}{\mu^2}+ \ldots \} \mbox{.}
\end{eqnarray*}
The linearized equations of motion are thus found to be
\begin{eqnarray*}
&\delta& m^2(q)= \lambda \varphi_0 \delta {\bar \varphi}(q) + 
\frac{\lambda}{2} \Pi_0(q^2) \delta m^2 (q) \mbox{,} \\
&(&-q^2 + \mu^2) \delta {\bar \varphi}(q) + 
\frac{\lambda}{2} \varphi_0 \Pi_0(q^2) \delta m^2 (q)=0 \mbox{.}
\end{eqnarray*}
These equations are equivalent to those derived by Kerman and Lin \cite{LIN}
in the functional Schr\"odinger picture. A compact and elegant construction
is provided by our manifestly covariant formalism.

When the momentum cutoff $\Lambda$ goes to $\infty$ the quantity $\lambda$
$\Pi_0$ has a finite limit while the product $\lambda$ $\varphi_0$ goes to
zero and the first equation becomes
\begin{displaymath}
\left[ 1- \frac{\lambda}{2} \Pi_0(q^2) \right] \delta m^2 (q)=0.
\end{displaymath}
The linearized evolution equations have two types of solutions.
The first type of solutions are given by  
\begin{displaymath}
\delta m^2 = 0, ~~~~\delta \varphi (q) = f(q) \delta (q^2 - \mu^2) \mbox{,}
\end{displaymath}
where $\mu$ is the self consistent solution of the gap equation
(\ref{GAP}) and $f(q)$ is a regular function of $q = (q_0, q_1, q_2,
q_3)$ to be specified later by boundary conditions.  In the second type
\begin{displaymath}
\delta m^2 (q) =~(q^2- \mu^2) \delta \varphi~~~~ \delta \varphi= 
f(q) \delta (q^2 - M^2 ) \mbox{,}
\end{displaymath}
One important difference with the solutions of the first type is that now 
$M^2$ is a solution of the equation
\begin{displaymath}
1- \frac{\lambda}{2} \Pi_0(M^2)=0.
\end{displaymath}
This equation is just the lowest order Bethe-Salpeter equation for the
$\varphi^4$ field theory.  For small enough values of the renormalized
coupling it has a single solution $M^2$ i.e. a single bound state
\cite{LIN}.

To construct Lorentz boost invariant solutions inside the forward
light cone ($x_0^2 - x_3^2 = t^2 - z^2 > 0,~~ x_0 = t > 0$), it is
convenient to introduce the light cone coordinates by
\begin{eqnarray*}
x_0 & = & \tau \cosh \eta, ~~~~~~  x_3 = \tau \sinh \eta, \mbox{,}  \\
q_0 & = & \sigma \cosh y, ~~~~~~  q_3 = \sigma \sinh y \mbox{,}
\end{eqnarray*} 
where $ \sigma = \pm \sqrt{ q_0^2 - q_3^2} $ and $y$ is the usual
rapidity variable.  Then the Fourier integral for $\delta \varphi$
for the first type of solutions is written as
\begin{eqnarray*}
\delta \varphi (x) & = &   
\frac{1}{(2 \pi)^4} \int_{-\infty}^{\infty} d \sigma \sigma 
\int_{-\infty}^{\infty} dy \int d {\bf q}_{\perp} 
f (q) \delta ( q^2 - \mu^2 ) \\ & & \quad \times \exp \left[ i \sigma 
\tau \cosh (y - \eta) - i {\bf q}_{\perp} \cdot {\bf x}_{\perp} \right] 
\mbox{,}  
\end{eqnarray*}
where ${\bf x}_{\perp} = (x_1, x_2)$ and ${\bf q}_{\perp} = (q_1,
q_2)$.  We seek the form of $f(q)$ which gives $\varphi (x)$
independent of $\eta$ and ${\bf x}_{\perp}$.  By noting that $q^2 =
\sigma^2 - {\bf q}_{\perp}^2$ does not depend on $y$, one can show by
inspection that the required form is given by
\begin{displaymath}
f (q) = \delta^{(2)} ( {\bf q}_{\perp} ) g (\sigma ) \nonumber
\end{displaymath}
Indeed, the Fourier integral can now be performed explicitly and the
result can be written by a linear combination of the Bessel function
and the Neumann function of order zero, $J_0$ and $N_0$:
\begin{displaymath}
{\bar \varphi} (x) = \varphi_0 + 
\alpha J_0 (\mu \tau) + \beta N_0 (\mu \tau) ,
\end{displaymath}
where we have used the following integral expressions for $J_0$ and
$N_0$ \cite{ABRAMOWITZ}:
\begin{eqnarray*}
J_0 (x) & = & \frac{2}{\pi} \int_0^{\infty} dt \sin ( x \cosh t ) \\
N_0 (x) & = & - \frac{2}{\pi} \int_0^{\infty} dt \cos ( x \cosh t ) 
\end{eqnarray*}
The two coefficients $\alpha$ and $\beta$ are given by the value of
the function $g(\sigma )$ at $\sigma = \pm \mu$ 
\begin{displaymath}
\alpha = \frac{1}{32\pi^3} \left[ g(-\mu) - g(\mu) \right], ~~~
\beta = \frac{i}{32\pi^3} \left[ g(\mu) + g(-\mu) \right] 
\end{displaymath}
These two coefficients may be alternatively given by the initial
conditions of $\varphi (\tau)$ and its derivative at $\tau = \tau_0$
when the evolution of the meson field starts.

The appearance of the Bessel (and Neumann) function may be easily
understood if we rewrite the Klein-Gordon equation 
$\left( \Box + \mu^2 \right) \varphi (x) = 0$ using the light cone
variables \cite{COOPER}:
\begin{displaymath}
\left[\frac{\partial^2}{\partial \tau^2}
+ \frac{1}{\tau}\frac{\partial}{\partial \tau} - 
\frac{1}{\tau^2}\frac{\partial^2}{\partial \eta^2} 
- \frac{\partial^2}{\partial x_{\perp}^2} + \mu^2
\right] \varphi (x) = 0 \mbox{.}
\end{displaymath}
For $\varphi (x)$ independent of $\eta$ and ${\bf x}_{\perp}$, this
equation becomes a form of Bessel's differential equations whose two
independent (real) solutions are given by $J_0$ and $N_0$.

Boost invariant solutions of the second type can be obtained in a
similar way:
\begin{eqnarray*}
&{\bar \varphi}& (x) = {\bar \varphi}_0 + 
\alpha J_0 (M \tau) + \beta N_0 (M \tau ) , \\
&m&^2 (x) = \mu^2+ ( M^2 - \mu^2 ) 
\left[ \alpha J_0 (M \tau) + \beta N_0 (M \tau) \right] . 
\end{eqnarray*}
We thus learn that for large values of the proper time the system
returns, as postulated above, to the vacuum state for both types of
solutions, which justifies a posteriori the approximation scheme we
have developped.

\paragraph*{Discussion. ---}
\label{sect_discussion}

The first type of solution leads to a momentum distribution of the
form $1/ \sqrt{k^2 + \mu^2}$ which is time independent and boost
invariant. It describes the Lorentz invariant expanding coherent state
of a meson gas. The second type of solution involves the solution of
the lowest order Bethe-Salpeter equation and can thus be interpreted
as an expanding coherent state of two-meson bound states.  Solutions
exist only for a negative bare coupling constant, as discussed in
reference
\cite{LIN}.

Although our formulae were derived for the specific case of a
single scalar field with pure Gaussian states and a particular
renormalization scheme involving a negative bare coupling,
they exhibit general features
of boost invariant boundary conditions which are expected to
hold for other more general models. In this respect it is instructive to
compare our formulae with the results obtained numerically
by Cooper {\it et al} \cite{COOPER}. These calculations concern
the case of a sigma model with statistical mixtures
as initial conditions.
They are expected to be relevant to discuss the
possible formation of a chiral condensate in ultra relativistic collisions.
They use a {\it finite} momentum cutoff and a {\it
positive} bare coupling constant, a case to which our formulae are not a priori
immediately applicable. There is nevertheless a striking agreement between
our formulae involving Bessel functions of order zero and
figure 5 of Cooper {\it et al} which shows the variation of the square mass
$\mu^2$ + $\delta m^2$ as a function of the proper time $\tau$. It is
interesting to note that the results of Cooper {\it et al} are well
described by a {\it single} Bessel 
function whereas our previous analysis suggests
a superposition of two Bessel functions with characteristic scales $\mu$ and
$M$. One should remember 
however that since Cooper {\it et al} have a positive coupling constant
there are no two- meson bound states. This explains why the scale $M$ is
absent but still leaves one unanswered question. Indeed without
bound states our formulae predict no variation at all in the square mass.
The answer to this question is that our derivation was made in the case
of an infinite momentum cutoff whereas  Cooper {\it et al} use a finite one.
This difference produces the non vanishing value of $\delta m^2$.

By looking at the results of Cooper {\it et al} it is also worthwhile
to note that some of the initial conditions make the small amplitude
approximation valid for the whole evolution.  Our formulae also
explain nicely why increasing the formation time leads to larger
asymptotic oscillations as a result of the $1/ \sqrt{\tau}$ behavior
of the Bessel functions. They do not explain why there are in this
case less instabilities since this involves the small proper time
domain in which our approximation scheme is expected to break
down. Our analytic asymptotic expressions appear nevertheless as a
useful reference to analyze some of the physics of the boost invariant
expansion of a meson gas. Results using the sigma model with a finite
cutoff will be presented in a forthcoming publication.

\paragraph*{Acknowledgments. ---}
\label{sect_ack}
One of us (D. V.) wishes to thank Prof. Yosuke Nagaoka and the members
of the Yukawa Institute for Theoretical Physics of the Kyoto
University for the hospitality extended to him during the summer of
1995. He also wishes to express his appreciation to the Japan Society
for the Promotion of Science for the attribution of a JSPS fellowship
to visit the Yukawa Institute in the spring of 1996. Stimulating
discussions with Hagop Sazdjian on the properties of the
Bethe-Salpeter equation for a scalar field are gratefully
acknowledged.  The work of T. M. has been supported in part by the
Grant-in-Aid for Scientific Research \# 06640394 of Ministry of 
Education, Science, and Culture of Japan.

\end{document}